# MSE FiTS: the ultimate multi-fiber optic transmission system

Kim Venn*[a], Darren Erickson[b], David Crampton[b], Rafal Pawluczyk[c], Paul Fournier[c], Pat Hall[d], Colin Bradley[a], Alan McConnachie[b,e], John Pazder[b], Farbod Jahandar[a], Stephanie Monty[a], Jooyoung Lee[a], Celine Mazoukh[c], Collin Kielty[a], Victor Nicolov[a], Kei Szeto[e], Alexis Hill[e]

[a]University of Victoria, 3800 Finnerty Rd., Victoria BC, Canada V8P 5C2
[b]National Research Council Canada, Herzberg Astronomy and Astrophysics Research Centre, 5071 W. Saanich Rd, Victoria BC, Canada V9E 2E7
[c]FiberTech Optica Inc., 330 Gage Avenue Suite 1, Kitchener ON, Canada N2M 5C6
[d]York University, 4700 Keele St., Toronto ON, Canada M3J 1P3
[e]Maunakea Spectroscopic Explorer, 65-1238 Mamalahoa Hwy, Kamuela HI, USA 96743

## ABSTRACT

The Maunakea Spectroscopic Explorer (MSE) is a next-generation observatory, designed to provide highly multiplexed, multi-object spectroscopy over a wide field of view. The observatory will consist of (1) a telescope with an 11.25 m aperture, (2) a 1.5 square-degree science field of view, (3) fibre optic positioning and transmission systems, and (4) a suite of low (R=3000), moderate (R=6000) and high resolution (R=40,000) spectrographs. The Fibre Transmission System (FiTS) consists of 4332 optical fibres, designed to transmit the light from the telescope prime focus to the dedicated spectrographs. The ambitious science goals of MSE require the Fibre Transmission System to deliver performance well beyond the current state of the art for multi-fibre systems, e.g., the sensitivity to observe magnitude 24 objects (@ SNR=2) over a very broad wavelength range (0.37 – 1.8 μm) while achieving relative spectrophotometric accuracy of < 3% and radial velocity precision of 20 km/s (@ SNR=5).

This paper details the design of the FiTS fibre system. It places FiTS into context with existing and planned spectroscopic facilities, such as Subaru/PFS, KPNO/DESI, ESO/4MOST and Gemini/GRACES. The results and lessons learned from GRACES are particularly applicable, since FiTS and GRACES share many team members, including industrial partner FiberTech Optica (Kitchener, ON). The FiTS system consists of 57 identical fibre cables. These cables have been designed to be modular, facilitating efficient construction and automated acceptance testing. *Each cable* consists of 76 fibres, including 57 fibres feeding light to the low and moderate resolution spectrographs and 19 fibres feeding the high-resolution spectrographs. Thus, the MSE/FiTS consists of 4332 fibres in total. Novel construction techniques utilizing continuous high-NA (*f*/2) fibres, pioneered by FiberTech Optica, are outlined and test results showing < 5% focal ratio degradation (FRD) in V-band are presented. The effect on FRD from varying the input *f*/# is also shown. Where test data is unavailable, system error budgets have been created to assess design choices on options such as fibre material, anti-reflection coatings, and fibre-optic connectors. The planned facility for automated optical and mechanical testing of the cables is also presented.

*kvenn@uvic.ca; phone 1 (250) 472-5182

# 1. OVERVIEW OF MSE OBSERVATORY

The Maunakea Spectroscopic Explorer (MSE) facility is an 11.25 m upgrade of the 3.6 m Canada France Hawaii telescope on Maunakea, Hawaii. MSE will have a wide field of view (1.5 square degrees) optimized for multi-object spectroscopy at several different spectral resolutions from R = 3000 to 40,000, and over a broad wavelength range from 0.37 to 1.8 μm[1][2]. MSE is designed as a survey facility, including millions of faint target (e.g., V = 24 in 3600 s at R=3000), spread over tens of thousands of square degrees. It will enable transformational science in several fields of astrophysical research, such as; the characterization of exoplanet hosts, the reconstruction of the Milky Way Galaxy through chemical tagging and precision kinematics of its stars, the search for the effects of dark matter on stellar streams, the environmental influences on galaxy formation since 'cosmic noon', and measuring black hole masses through reverberation mapping of quasars. The MSE will also provide invaluable support for other key astrophysical surveys and facilities, such as Gaia, LSST, SKA, TMT, and more[1][2].

# 2. MSE SCIENCE REQUIREMENTS

It is the goal of MSE to provide the ability to observe more and fainter targets, with higher precision and higher cadence or baseline, than other multi-object facilities. This paper describes one of the key components of MSE, the Fiber Transmission System (FiTS), which takes the light emanating from the fiber positioners to the spectrographs. To achieve the scientific goals, MSE must obtain at least 3200 spectra per exposure at low and moderate (LMR) spectral resolutions and 1000 spectra per exposure at high resolution (HR). Excellent throughput is required from 370 nm to the J band, and potentially the H band, at low and moderate resolution, and to 900 nm at high resolution. The requirement on the loss in the fibers due to focal ratio degradation (FRD) is <5%, a challenging but obtainable requirement. Other challenging requirements for FiTS include radial velocity precision of 100 m/s (HR mode) and 0.1% sky subtraction[1][2]. These specifications demand high throughput of all FiTS fibers, and control of variations in the FRD, as well any other effects that fibers are susceptible to such as the distribution of the far field intensity. Both spectral and flat calibrations taken during daytime (and at night if necessary) will be needed to reach these requirements. Thus, FiTS is required to have very high stability and repeatability over (at least) a 24 h period.

# 3. EXISTING FIBRE TRANSMISSION FACILITIES

Optical fibres have been employed for wide field spectroscopic surveys for many years, particularly for large redshift surveys (e.g., 2DF and subsequent surveys). In general, the throughput and stability achieved through fibre transmission systems has not been as good as through multi-slit systems and consequently most very faint surveys have used the latter, at low spectral resolution. Now, however there is increasing desire to observe faint targets with moderate spectral resolution over very large areas of the sky and consequently fibre systems are the only practical solution. Some excellent information on current and proposed projects was recently discussed at the "Multiplexed Fiber Spectroscopy for Keck & TMT" meeting[3]. Here we summarize aspects of some of the most recent projects to illustrate the scientific drivers and engineering challenges.

### 3.1 Subaru PFS

The Subaru 8.2 m telescope stands out for its wide field capability. The Subaru Prime Focus Spectrograph (PFS) will use 2400 phi-theta fiber positioners in a 1.3 deg prime focus field to feed four relatively low resolution spectrographs for cosmological surveys and to study galactic evolution[4]. The fibers are fed by the wide field corrector at f/2.8 and are 127 μm in diameter or 1.1" on the sky. The spectrographs will cover the entire 380 nm – 1260 nm range and the fiber spacing is 224 μm at the slit. Since Subaru is a multipurpose telescope, PFS will be frequently taken on and off the telescope necessitating two connectors in the 50 m long fiber cables, which add to the challenges of maintaining performance. Scientific operation is expected to begin in 2021.

### 3.2 DESI

DESI, the Dark Energy Spectroscopic Instrument, is a project to install a multi-fiber spectroscopic facility on the KPNO 4 m telescope that will measure the redshifts of 25 million galaxies to probe the nature of dark energy and test General Relativity[5]. 5000 phi-theta positioners cover a 7.5 square degree (or 3.2 deg, diameter) field. The fibers are fed at f/3.9 and the core diameter of the fiber is 107 μm. Optical fibers run 49.5 m from each positioner to the spectrographs. The fiber system throughput is projected to vary from 0.5 at 360 nm to 0.8 at 980 nm. Completion expected in 2020.

### 3.3 4MOST

4MOST is a fiber fed wide field spectroscopic facility on the 4 m VISTA telescope in Chile, intended to complement a number of large-area, European space-based observatories (Gaia, eROSITA, EUCLID, PLATO), and future ground-based, wide-area survey facilities (e.g., LSST, ALMA, SKA)[6]. It will have 2400 fiber positioners (f/3.3) of the tilting spine type in a 4 square degree field that will feed both low resolution (R~5000) and high resolution (R~20,000) spectrographs, covering the 370 – 950 nm wavelength range. The 85 μm fibers will subtend 1.4" on the sky. The fibers that feed the LR spectrographs will be ~15 m, and the HR fibers will be ~20 m, but all fibers will pass through connectors to facilitate maintenance. Completion expected in 2021.

### 3.4 GRACES

Our MSE FiTS project builds upon a collaboration between NRC Canada and FiberTech Optica (FTO) that produced the extremely successful GRACES (Gemini Remote Access Echelle Spectrograph) fiber system. GRACES is a fibre optic link between the Gemini North Observatory and the Espadons spectrograph located in the CFHT dome[7]. Despite the 270 m length of the fibres, the average loss due to FRD is less than 14%, and the internal transmittance is better than 83% at 800 nm making the spectrograph more efficient than Keck HIRES at wavelengths > 600 nm in spite of 100/64 telescope aperture advantage. The fiber link has proved to be extremely stable.

## 4. MSE FIBRE TRANSMISSION SYSTEM (FITS)

### 4.1 The MSE Field

Fibres are placed on objects within the MSE field using the Sphinx positioning system, supplied by the Australian Astronomical Observatory (AAO)[8]. This fibre positioner uses 4332 individual tilting spines, with one fibre per spine. The positioning system is divided into 3 identical sectors. Within each sector are 19 Sphinx Modules. Each module contains two rows of spines with 57 fibres feeding the Low/Moderate Resolution (LMR) spectrographs and 19 fibres feeding the High Resolution (HR) spectrographs (see Fig. 1). The patrol radius of each spine is sufficiently large to allow complete field coverage for both the LMR and HR spectrographs.

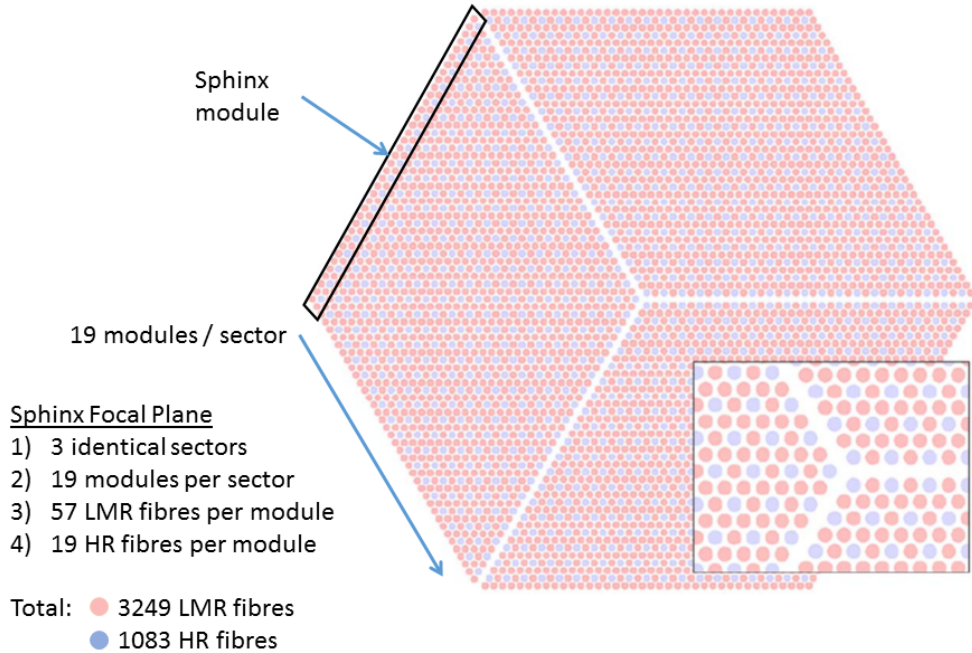

Figure 1: Schematic of the MSE field showing distribution of LMR fibres (pink) and HR fibres (blue)

### 4.2 FiTS Systems Design

The Fibre Transmission System is required to transmit light from each fibre positioner at the prime focus of the telescope to two types of spectrographs (see Fig. 2). Six identical Low/Moderate Resolution spectrographs cover a wavelength range of 0.37 – 1.8 μm with a spectral resolution of R=3000 and R=6000 [9][10]. Two identical High Resolution spectrographs cover a wavelength range of 0.37 – 0.9 μm with a spectral resolution of R=40,000 [9][11]. The location of these spectrographs is the subject of an ongoing observatory trade study. For the purpose of this design, we are assuming the two HR spectrographs are located on the instrument platforms of the telescope and the six LMR spectrographs are located in the lower Coudé room. This arrangement is opposite from the MSE project conceptual design configuration. However, this arrangement will maximize throughput in the HR instruments due to the shorter cable length from prime focus to the instrument platforms.

Light enters FiTS at the Prime Focus Assembly, a structural component used to interface the fibre cables to the telescope Top End Assembly and includes interfaces to the Positioning System (PosS) and the Instrument Rotator (InRo)[12]. From this location, the HR cables are routed along two unique paths across the telescope structure and pass through the centre of the elevation bearings before entering the two HR spectrographs. The total path length of the HR fibres is approximately 35 m. All of the LMR cables are routed along a separate path and pass through a 180° service loop near the azimuth platform before passing through a hole in this platform and entering the LMR spectrographs in the lower Coudé room. The total path length of the LMR fibres is approximately 50 m. This is shown schematically in Fig. 3.

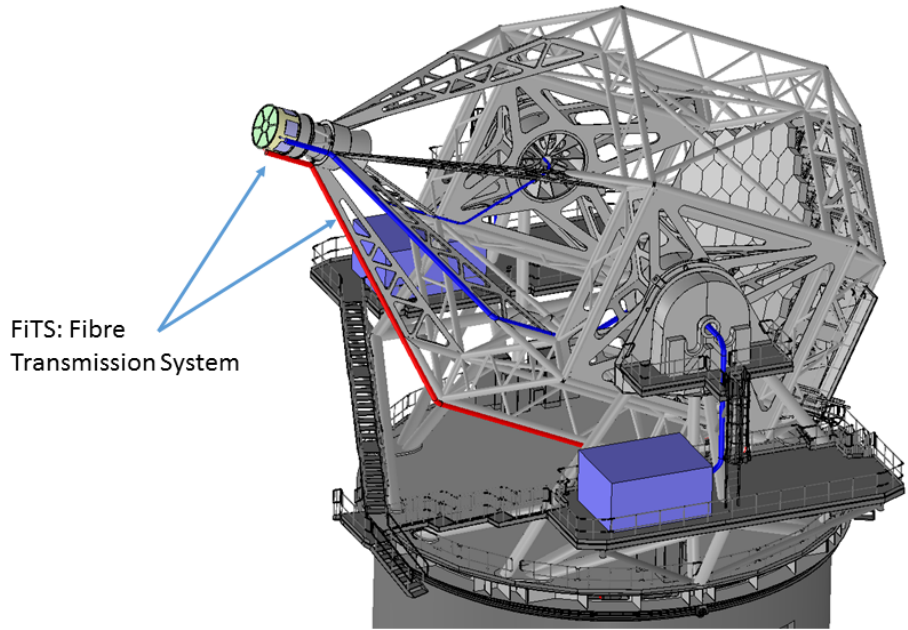

**Figure 2: Location of FiTS Prime Focus Assembly**

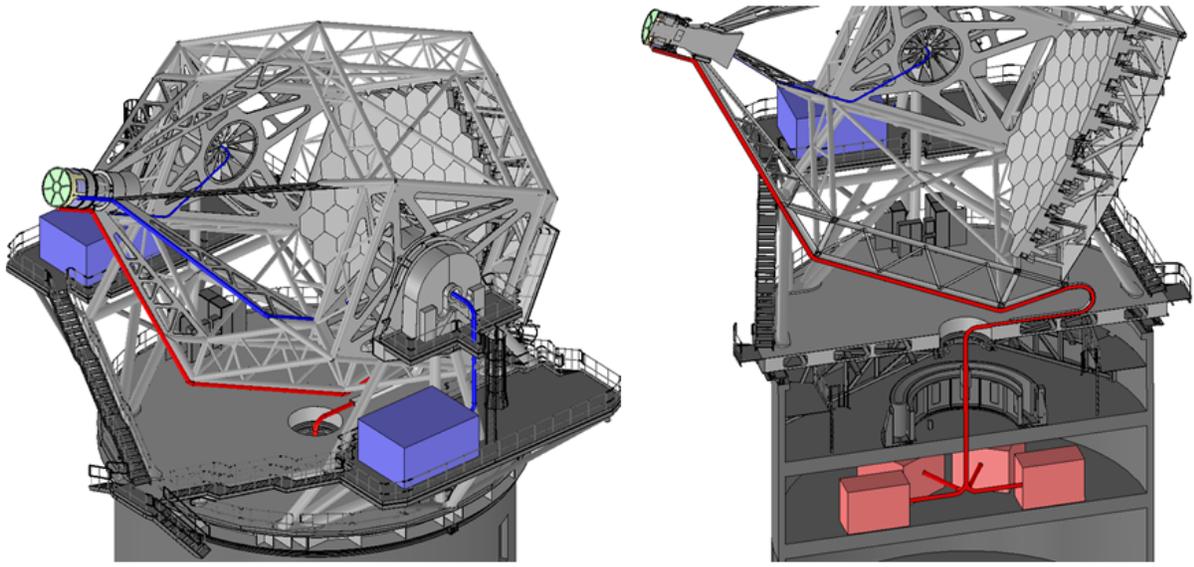

**Figure 3: FiTS cable routing. HR fibres (blue lines, left image). LMR fibres (red line, right image).**

### 4.3 Fibre Cable Construction

The fibre cables were designed to be modular to facilitate design, fabrication, testing and maintenance. Following this philosophy, each Sphinx Positioner module has a corresponding fibre cable. This results in 57 identical cables, each with one branch feeding a LMR spectrograph and one branch feeding a HR spectrograph. This is shown schematically in Fig. 4.

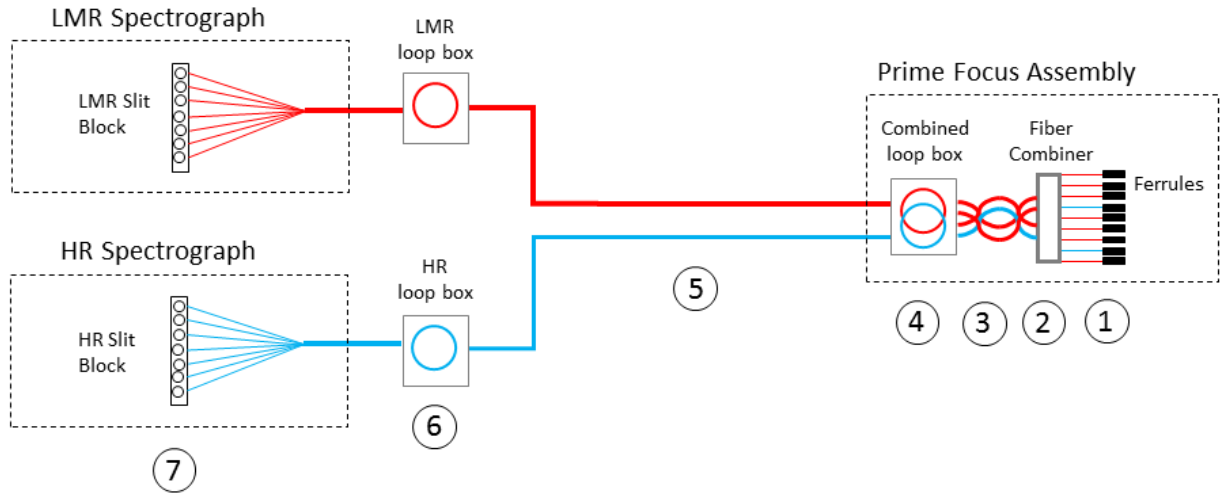

**Figure 4: Cable Schematic.** Photons enter the Prime Focus Assembly at right. See below for descriptions of components 1 to 7

The fibre cable can be described by following light through the system (numbered items below correspond to Fig. 4). Further details are given in the following sections.
1) The input end of each fibre is bonded within a tubular ceramic ferrule. This provides a stiff interface to the positioning system spines (not shown). There is one ferrule for each fibre.
2) Individual fibres are then grouped into 4 protection tubes at the fibre combiner. Three tubes contain 19 LMR fibres each (57 LMR fibres total), while a fourth tube contains 19 HR fibres. These 4 tubes exit the back of each fibre combiner.
3) The protection tubes are wound through a helical section to provide rotation compliance to the positioners, ferrules, and fibre combiners, as they rotate with the field.
4) Fibre tubes enter loop boxes at the fixed end of the helical section. Loop boxes provide access to bare fibres for splices and repairs if necessary.
5) Fibre cables route across the telescope structure (57 fibres + spares in the LMR cable tube, 19 fibres + spares in the HR cable tube).
6) A second set of loop boxes are located near the spectrographs, providing access to bare fibre for repairs if necessary.
7) Individual fibres are arranged into a one-dimensional array using slit blocks. The slit blocks act as the entrance slit to the spectrographs.

An important aspect to this design is that the fibres are continuous, with no breaks or connectors. This choice poses some potential difficulties for integration, but results in significantly higher throughput and FRD stability, which are key science requirements. It also simplifies the cable construction process by removing the need to cleave, polish, and install connectors or fusion splice more than 4332 individual fibres. This decision remains part of an on-going system trade and will be revisited as we work with other MSE sub-systems to develop an integration and testing plan.

A second key requirement addressed by the cable design is loss due to FRD, which causes the output beam to have a faster f/ratio than the input, potentially overfilling the spectrograph collimator[13]. FRD is known to increase as fibres are stressed or manipulated. Testing at FiberTech Optica and the

University of Victoria indicate that high Numerical Aperture (NA) fibres suffer from less FRD than low NA fibres[14]. Candidate fibres for FiTS include the Polymicro FBP and CeramOptec WF series. Both fibres have NA = 0.28, which can accept the *f*/2.0 beam from the telescope prime focus optics. FiberTech Optica also has proprietary cable materials and production techniques for cables that are largely insensitive to mechanical motions. A cable test facility is in development at the University of Victoria to test throughput and FRD under realistic conditions and a variety of telescope motions (see Section 5.2).

**4.4 Prime Focus Assembly**

The ferrules, fibre combiners, helical tubes and loop boxes are contained within the Prime Focus Assembly (see Fig. 5). This structure has two main functions; (1) to organize the multitude of individual fibres into a smaller number of multi-fibre cables, and (2) to accommodate the motion required by the rotating field.

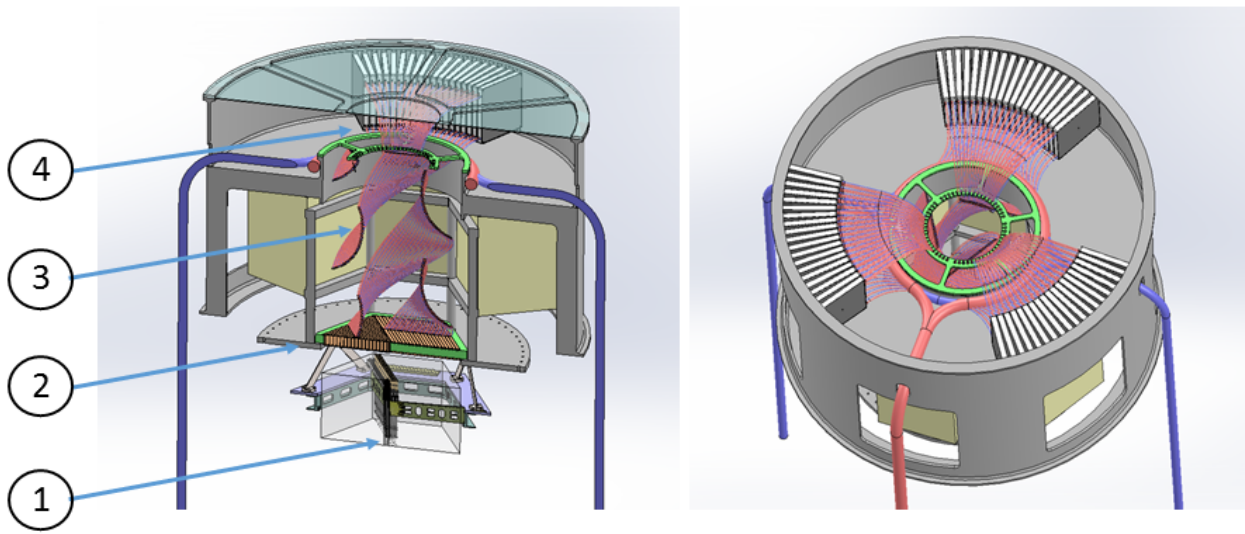

**Figure 5: FiTS Prime Focus Assembly showing: 1) Positioning system (simplified), 2) fibre combiner, 3) helical tubes, 4) loop boxes**

Individual fibres pass from the spines of the positioning system to the fibre combiner. Within the fibre combiner, fibres are grouped together into polymer protection tubes (see Fig. 6). The protection tubes are wound through a helical section to provide rotation compliance to the positioners, ferrules, and fibre combiners, as they rotate with the field.

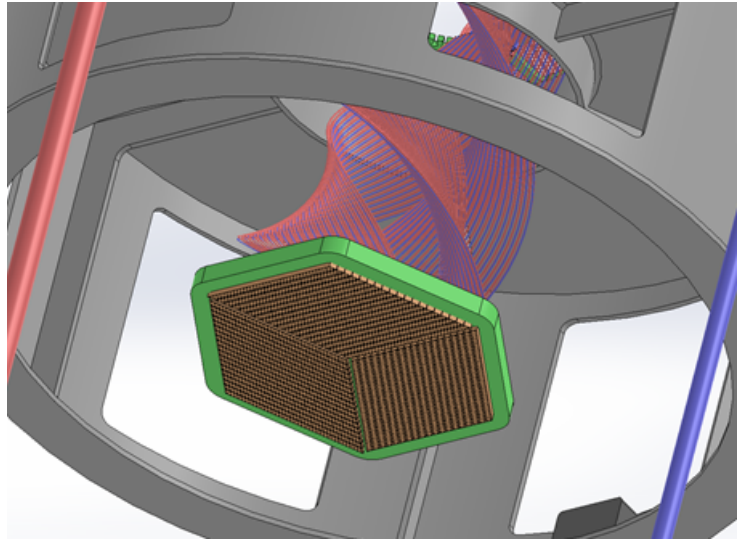

**Figure 6: Fibre Combiner and Helical Tubes**

Fibre tubes enter loop boxes at the fixed end of the helical section (see Fig. 7). Fibres are bare (not contained within tubes) inside the loop boxes, thereby giving access to make fusion splices for assembly or repair. The fibres make approximately 1.5 revolutions inside the box, requiring a meter of length. All fibres exit the loop box within one of 2 additional protection tubes. One tube contains 57 LMR fibres, the other contains the 19 HR fibres. Extra fibres may also be included as spares to facilitate future repairs.

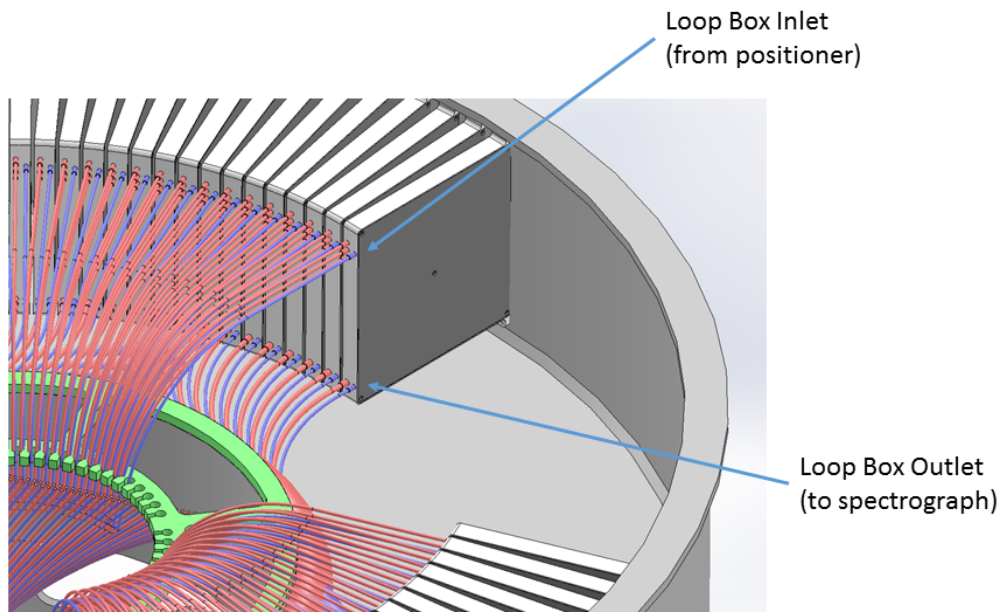

**Figure 7: Combined Loop Boxes**

### 4.5 Cable Routing and Management

Once the fibre cables leave the Prime Focus Assembly, they traverse the telescope top end prime focus support spiders and the telescope structure en-route to the spectrographs. The cables are attached to brackets which keep them organized while allowing a single cable to be removed for repair as needed (see Fig. 8). The brackets also minimize the apparent width of the cable group to remain within the shadow of the support spiders thereby not causing any additional vignetting of the telescope primary mirror.

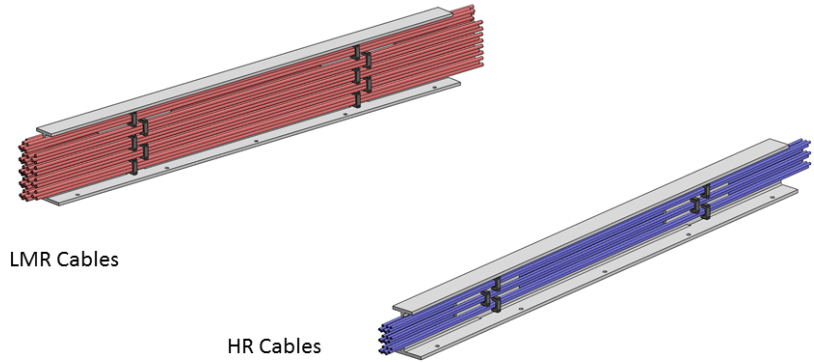

**Figure 8: Example sections of LMR and HR cable groups with support brackets**

### 4.6 Slit Blocks

The fibre cables terminate at the spectrographs in slit blocks. These blocks organize the fibres into a single row using precisely formed V-grooves (see Fig. 9). The spacing of the fibres is determined largely by the spectrograph design, with the lower limit set by the allowed cross-talk between adjacent spectra on the detector and the upper limit set by the overall slit length. The spacing is currently assumed to be 2x the fibre core diameter. The fibres from a single fibre cable form a small linear array (or slitlet). Slitlets are mounted to a common base plate, forming the complete entrance slit at the spectrograph.

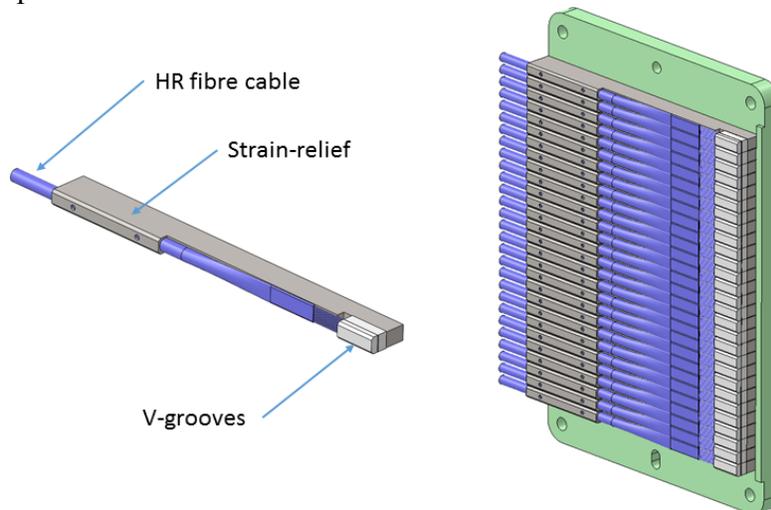

**Figure 9: Slitlet formed using V-grooves with a single fibre cable (left), spectrograph entrance slit formed by aligning multiple slitlets mounted to a common base plate (right).**

**4.7 Throughput Budgets and Modelling**

Optical throughput budgets have been created for the LMR fibres (see Fig.10) and HR fibres (see Fig. 11)[15], which include transmission losses through the fibre, reflection (Fresnel) losses at the input and output surfaces of the fibres, and coupling losses at the spectrograph due to focal ratio degradation. Transmission losses are given in the manufacturer's data sheet and are wavelength dependent. Fresnel losses were determined assuming bare fibre ends without anti-reflection coatings (3.5% per surface). FRD losses were estimated at 5%, based on the MSE science requirements.

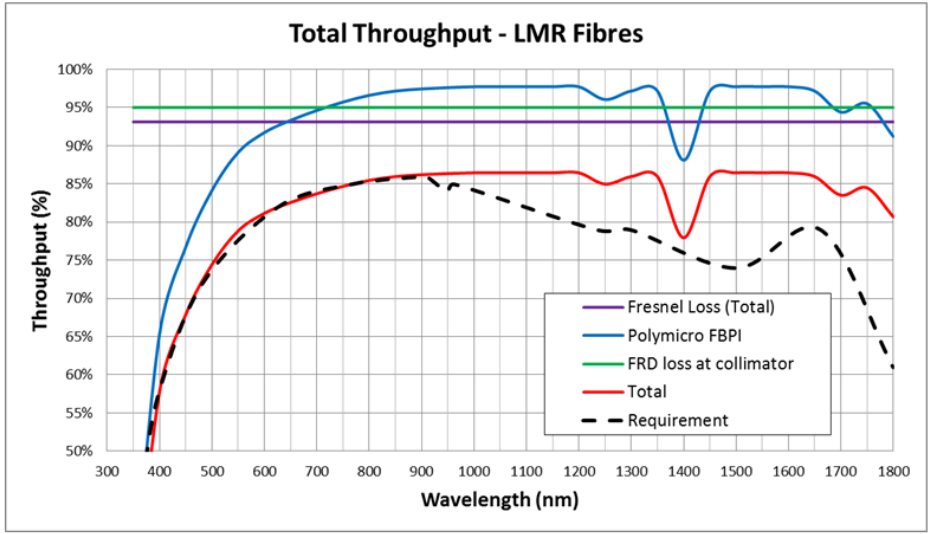

Figure 10: Polymicro FBPI (50 m) fibre throughput (LMR fibres): expected performance (red line), requirement (black dashed line)

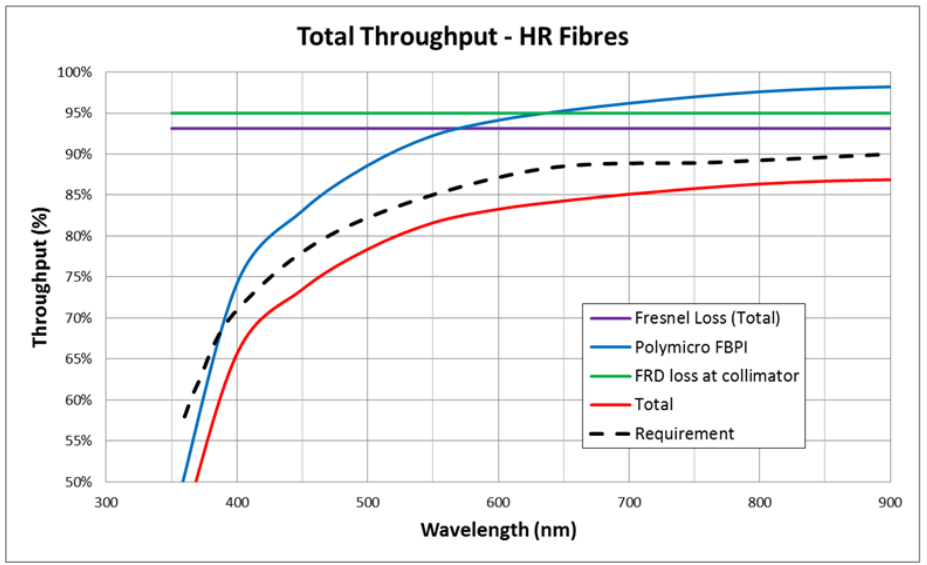

Figure 11: Polymicro FBPI (35 m) fibre throughput (HR fibres): expected performance (red line), requirement (black dashed line)

The expected performance of the LMR fibres meets the requirements in the bandpass from 370 – 900 nm, and exceeds the requirements from 900 nm to 1.8 µm. This is encouraging given the long fibre length (50 m) to the lower Coudé room, and the lack of AR coatings on the ends. On the other hand, the HR fibres are 3-5% below the requirement over most of the bandpass, and approximately 10-15% below at the bluest end. It is clear that these fibres would benefit from AR coating, and work is under way to identify a suitable coating. This result underscores the importance of locating the HR spectrographs on the instrument platforms, to minimize the overall length of fibre.

## 5. FABRICATION, INTEGRATION AND TESTING

### 5.1 Fabrication

Cable manufacturing will be performed by our industrial partner, FiberTech Optica, using their proven processes and materials. The production steps are generally as follows:
1. Lengths of LMR and HR fibres are pulled through the long sections of protection tubing (item 5, Fig. 4). This length of fibre will include a TBD number of spare fibres for repair purposes.
2. Components at the input end are then integrated
    a. The combined loop box is attached to the LMR and HR tubes, and the bare fibres are routed through the loop box (item 4, Fig. 4).
    b. Additional protection tubes are applied over the fibres in the helical section (item 3, Fig. 4).
    c. The fibre combiner is integrated onto the protection tubes, and individual fibres are routed through the combiner (item 2, Fig. 4).
    d. The individual fibres are cleaved and bonded into ceramic ferrules (item 1, Fig. 4).
3. Finally, the components at the output end are integrated
    a. The LMR and HR loop boxes are attached to the LMR and HR tubes, and the bare fibres are routed through the loop boxes (item 6, Fig. 4). Additional protection tubes are then applied over the fibres between the loop boxes and slit blocks.
    b. The individual fibres are cleaved and aligned into the slit blocks and bonded in place using a cover.

A single cable is estimated to take several weeks to fabricate, but multiple cables can be built in parallel. For example, once the fibres have been pulled through the long lengths of the protection tubes, and loop boxes integrated onto the ends, this assembly can be passed to the next stage, while a second cable is started.

### 5.2 Cable Testing

Once the cables are fabricated, they will undergo significant end-to-end optical testing to measure the throughput and FRD of each fibre under realistic telescope conditions. To accomplish this, the University of Victoria is building a fully-automated opto-mechanical test facility [14]. This facility will include sources and detectors to measure the optical performance of the fibres with mechanisms to bend/twist the fibre cable to mimic the motion of the telescope. The result is a measurement of the static and kinematic performance, for an estimate of the fibre throughput stability. All tests will be performed under closed-loop computer control with minimal human intervention (except for installing and removing the cable). By fully automating this complicated metrology, the testing will be able to match the cable fabrication schedule.

Although the facility is currently under development, preliminary results are available. In Fig. 12, we show the output image of a ring test, used to quantify the FRD from a sample fibre provided by FTO. The image on the left is the full 2D illumination pattern while the image on the right is a slice through the central row of pixels used to measure the ring geometry. Post-processed results are shown in Fig. 13, indicating an FRD performance < 5% is possible when using high-NA (low $f/\#$) fibres. It is acknowledged that the ring test is primarily a qualitative measure of FRD performance, and future improvements to the test facility will include full-aperture FRD testing capabilities for the final fibre qualifications.

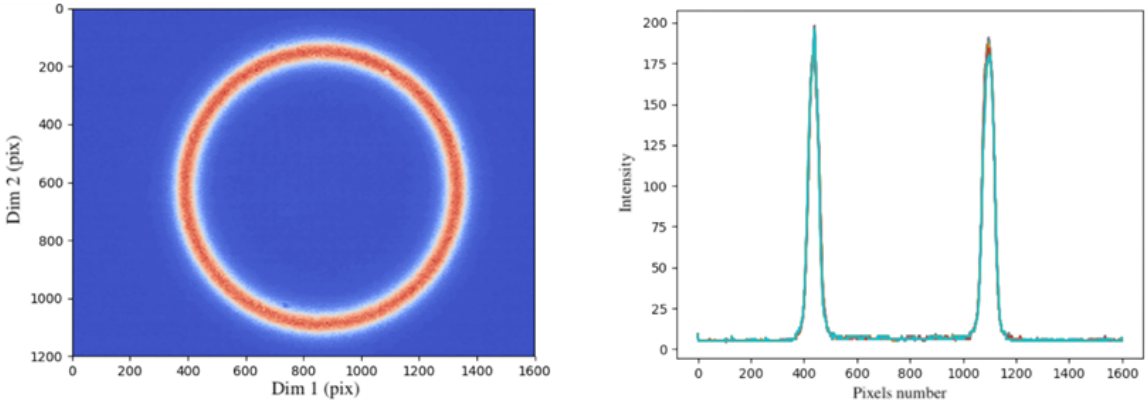

**Figure 12: Ring test image processing. Left panel shows the 2D illuminated ring pattern, while right panel shows a slice through the central row of pixels used to measure the ring geometry.**

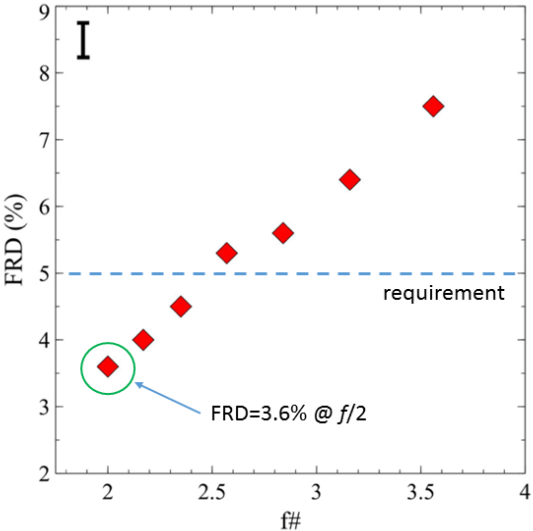

**Figure 13: FRD results from a candidate high-NA fibre using the ring test.**

# 6. CONCLUSIONS

The Fibre Transmission System is used to deliver light from up to 4332 objects in the MSE field to a suite of Low/Moderate and High Resolution spectrographs. To minimize focal ratio degradation and throughput losses, the system employs continuous lengths of high-NA fibres, operating at $f/2$. In addition to maximizing throughput, continuous fibres remove the need to cleave, polish, and install connectors, or fusion splice each fibre.

For the spectrograph locations stated in our study, initial analysis and testing indicates that the LMR fibres will meet their throughput and FRD requirements. This can be achieved even considering the long (50 m) length to the LMR spectrographs in the lower Coudé room and assumes no anti-reflection coating on the fibre ends. The HR fibre throughput budget shows performance that is 3-5% below the requirements (and 10-15% below at the bluest end of the bandpass). Future work will consider suitable AR coatings for these fibres to reduce the Fresnel losses, and may include the development of custom fibres with higher throughput.

The mechanical design of FiTS is relatively straightforward. These cables use proven construction techniques pioneered at FiberTech Optica to deliver performance that is stable in the presence of telescope motions. The Prime Focus Assembly provides the needed rotational compliance and organization of fibres into cables. Finally, the slit blocks create the required spectrograph entrance slit geometry, while maintaining the modular philosophy of the design. Challenges remain in meeting the stringent performance specifications, and in successfully integrating FiTS with the MSE telescope design and its sub-systems.

Cable production will require a significant amount of time, and necessitate parallel production techniques. To keep up with the required production rate, a fully automated opto-mechanical test facility is being developed. This facility will measure the FRD and throughput of every fibre in a cable, while the cable is subjected to bending and twisting motions to mimic the telescope movements.

# ACKNOWLEDGEMENTS

The authors and the MSE collaboration recognize and acknowledge the cultural importance of the summit of Maunakea to a broad cross section of the Native Hawaiian community. KV, CB, FJ, SM, JL and CK acknowledge the support of the NTCO CREATE program, NSERC funding reference number 498006-2017. PH acknowledges support from the Natural Sciences and Engineering Research Council of Canada (NSERC), funding reference number 2017-05983, and for sabbatical support from the National Research Council Canada.